\documentclass{article}
\usepackage{amsmath}
\usepackage{amsthm}
\usepackage{amssymb}
\usepackage{natbib}
\usepackage{graphicx}

\newcommand{\ownint}[4]{{\int_{#1}^{#2} \! #3 \, \mathrm{d}#4}}

\newcommand{\innprod}[2]{\left\langle#1,#2\right\rangle}

\newcommand {\W}{\mathcal{W}}
\newcommand {\E}{\mathbb{E}}
\renewcommand {\P}{\mathbb{P}}
\newcommand {\R}{\mathbb{R}}

\newcommand{\topt}[2]{{\mathbf t_{#1}^{#2}}}
\newcommand{\id}{\mathbf i}
\newcommand{\Tan}{\mathrm{Tan}}
\newcommand{\X}{\mathcal X}
\begin{document}

\title{Statistical Aspects of Wasserstein Distances}
\author{Victor M. Panaretos\footnote{Institut de Math\'ematiques, \'Ecole polytechnique f\'ed\'erale de Lausanne, Switzerland;  email: victor.panaretos@epfl.ch} and Yoav Zemel\footnote{Institut f\"ur Mathematische Stochastik, Georg--August--Universit\"at G\"ottingen, Germany; email:  yoav.zemel@mathematik.uni-goettingen.de}}
\date{June 8, 2018}
\maketitle


\begin{abstract}
Wasserstein distances are metrics on probability distributions inspired by the problem of optimal mass transportation.  Roughly speaking, they measure the minimal effort required to reconfigure the probability mass of one distribution in order to recover the other distribution.  They are ubiquitous in mathematics, with a long history that has seen them catalyse core developments in analysis, optimization, and probability. Beyond their intrinsic mathematical richness, they possess attractive features that make them a versatile tool for the statistician: they can be used to derive weak convergence and convergence of moments, and can be easily bounded; they are well-adapted to quantify a natural notion of perturbation of a probability distribution; and they seamlessly incorporate the geometry of the domain of the distributions in question, thus being useful for contrasting complex objects. Consequently, they frequently appear in the development of statistical theory and inferential methodology, and have recently become an object of inference in themselves. In this review, we provide a snapshot of the main concepts involved in Wasserstein distances and optimal transportation, and a succinct overview of some of their many statistical aspects.
\end{abstract}



\section{Introduction}
Wasserstein distances are metrics between probability distributions that are inspired by the problem of optimal transportation.  These distances (and the optimal transport problem) are ubiquitous in mathematics, notably in fluid mechanics, partial differential equations, optimisation, and, of course, probability theory and statistics.  In addition to their theoretical importance, they have provided a successful framework for the comparison of (at times complex) objects in fields of  application such as image retrieval \citep{rubner2000earth}, computer vision \citep{ni2009local}, pharmaceutical statistics \citep{munk1998nonparametric}, genomics \citep{bolstad2003comparison,evans2012phylogenetic}, economics \citep{gini1914di} and finance \citep{rachev2011probability}, to name but a few. Indeed, while their origins lie with Monge's (primarily mathematical) enquiry into how to optimally transport a pile of earth of a given volume into a pit of equal volume but potentially different shape, Kantorovich's modern reformulation, which catalysed the development of this rich theory, was inspired by the concrete problem of optimal resource allocation. Unsurprisingly, there is a vast literature on Wasserstein distances and optimal transportation, originally rooted primarily in analysis and probability, but later branching out to quantitative fields well beyond. In statistics, Wasserstein distances play a prominent role in theory and methodology, and more recently have become an object of inference in themselves. In his thousand-page book, \cite{villani2008optimal} writes that reviewing the optimal transport literature is a ``\emph{dauntingly difficult task}". And, if one focusses more narrowly on \emph{statistical} aspects of Wasserstein distances, it is still impossible to carry out a comprehensive review in the order of thirty pages. We thus restrict ourselves to a high level overview of some salient aspects and main concepts, admittedly influenced by our own perspective and interests, and apologise for the inevitable omissions.

\subsection{Overview}
Wasserstein distances appear in statistics in several ways. We delineate three broad categories of statistical use of these distances, according to which we will structure our review:

\begin{enumerate}

\item[(1)] Wasserstein distances and the associated notion of an optimal coupling are often exploited as a versatile tool in asymptotic theory, due to the topological structure they induce and their relatively easy majorisation, and Section~\ref{sec:tool} reviews some of their appealing features in that context.

\item[(2)] In other cases, Wasserstein distances are employed as a methodological tool, in order to carry out statistical inference, primarily involving structural models and goodness-of-fit testing. Section~\ref{sec:inference} describes key methods and results in this vein.

\item[(3)] Finally, a recent trend in functional data analysis is to consider the space of probability measures equipped with a Wasserstein distance as a sample/parameter space itself, a direction that is taken up in Section~\ref{sec:statWass}.

\end{enumerate}

In contexts such as (2) and (3), it is often important to carry out explicit computations related to the Wasserstein distance, and Section~\ref{sec:numerics} gives a brief overview on such numerical aspects. First, though, the next subsection reviews the basic definitions and relevant notions that we require throughout our review.

\subsection{Basic Notions}  The $p$-Wasserstein\footnote{Also known as \emph{Mallows' distance}, \emph{Earth mover's distance}, \emph{(Monge--)Kantorovich(--Rubinstein) distance} or \emph{Fr\'echet distance} (when $p=2$).  The terminology \emph{Wasserstein distance} became popular, mainly in Western literature, following \cite{dobrushin1970prescribing} who studied some of its topological properties and referred to an earlier work by Wasserstein.  See \citet[page 118]{villani2008optimal} and \citet[page 4]{bobkov2014one} for more details.} distance between probability measures $\mu$ and $\nu$ on $\R^d$ is defined as
\begin{equation}\label{prob_definition}
W_p(\mu,\nu)
=\underset{\tiny \begin{array}{c}X\sim\mu \\Y\sim\nu\end{array}}{\inf}\left(\E \|X - Y\|^p\right)^{1/p},
\qquad p\ge1,
\end{equation}
where the infimum is taken over all pairs of $d$-dimensional random vectors $X$ and $Y$ marginally distributed as $\mu$ and $\nu$, respectively (an obviously nonempty set, since one can always construct independent random variables with prescribed marginals). For convenience, we shall use both notations $W_p(X,Y)$ and $W_p(\mu,\nu)$ interchangeably, whenever $X\sim\mu$ and $Y\sim\nu$.  The distance is finite provided the $p$-th moments exist, $\E \|X\|^p+\E\|Y\|^p<\infty$, and this will be tacitly assumed in the sequel. The definition generalises to laws defined on much more general spaces:  if $(\X,\rho)$ is any complete and separable metric space, $W_p$ can be defined in the same way, with $\|X-Y\|$ replaced by the metric $\rho(X,Y)$.  In particular, this setup incorporates laws on infinite-dimensional function spaces such as $L^2[0,1]$.  For simplicity, we restrict to the setting where $\mathcal{X}$ is a normed vector space, employing the notation $(\X,\|\cdot\|)$ henceforth.

The optimisation problem defining the distance is typically referred to in the literature as \emph{optimal transport(ation)} or the \emph{Monge--Kantorovich} problem. When $X$ and $Y$ take values on the real line, their joint distribution is characterised by specifying their marginal distributions and a copula \citep{sklar1959fonctions}.  Since the marginals here are fixed to be the laws of $X$ and $Y$, the problem is to find a copula that couples $X$ and $Y$ together as ``tightly" as possible in an $L_p$-sense, on average; if $p=2$ then the copula one seeks is the one that maximises the correlation (or covariance) between $X$ and $Y$, i.e., the copula inducing maximal linear dependence.

The Wasserstein distances $W_p$ are proper distances in that they are nonnegative, symmetric in $X$ and $Y$, and satisfy the triangle inequality.  A compactness argument shows that the infimum in their definition is indeed attained (if $\X$ is complete).  The space of measures with $p$-th moments finite, the \emph{Wasserstein space} $\W_p(\X)$, when endowed with the distance $W_p$, is complete and separable if $\X$ is so.  Although many other metrics can be defined on the space of probability measures \citep{rachev1991probability,gibbs2002choosing}, Wasserstein distances exhibit some particularly attractive features:
\begin{itemize}
\item They incorporate the geometry of the ground space $\X$:  if $X$ and $Y$ are degenerate at points $x,y\in\X$, then $W_p(X,Y)$ is equal to the distance between $x$ and $y$ in $\X$.  This property hints at why Wasserstein distances are successful in imaging problems and why they can capture the human perception of whether images are similar or not (see Section~\ref{sec:statWass}).

\item Convergence of $X_n$ to $X$ in Wasserstein distance is equivalent to convergence in distribution, supplemented with $\E \|X_n\|^p\to \E\|X\|^p$.  This makes $W_p$ convenient for proving central limit theorem-type results (see Section~\ref{sec:tool}).

\item Since they are defined as the solution of minimisation problems, they are quite easy to bound from above:  \emph{any} joint distribution with the correct marginals provides an upper bound for the Wasserstein distance (see Section~\ref{sec:tool}).  Moreover, they enjoy some differentiability, allowing for application of the delta method (see Section~\ref{sec:inference}).
\end{itemize}

\noindent Further to the ``probabilistic" definition (Definition \ref{prob_definition}), one can consider the ``analytic" definition, which helps dissect the structure of the Monge--Kantorovich optimisation problem:
\begin{equation}\label{analyst_definition}
W_p(\mu,\nu)
=\left (  \inf_{\gamma\in\Gamma(\mu,\nu)}\ownint{\X\times\X}{}{\|x-y\|^p}{\gamma(x,y)}  \right)^{1/p}.
\end{equation}
Here $\Gamma(\mu,\nu)$ is the set of probability measures $\gamma$ on $\X\times\X$ satisfying $\gamma(A\times\X)=\mu(A)$ and $\gamma(\X\times B)=\nu(B)$ for all Borel subsets $A,B\subseteq\X$.  Elements $\gamma\in\Gamma(\mu,\nu)$ are called \emph{couplings} of $\mu$ and $\nu$, i.e., joint distributions on $\X\times\X$ with prescribed marginals $\mu$ and $\nu$ on each ``axis", which hopefully elucidates the equivalence to Definition \ref{prob_definition}.  Definition \ref{analyst_definition} has a simple intuitive interpretation in the discrete case: given a $\gamma\in \Gamma(\mu,\nu)$, and any pair of locations $(x,y)$, the value of $\gamma(x,y)$ tells us what proportion of $\mu$'s mass at $x$ ought to be transferred to $y$, in order to reconfigure $\mu$ into $\nu$. Quantifying the effort of moving a unit of mass from $x$ to $y$ by $\|x-y\|^p$ yields the interpretation of $W_p(\mu,\nu)$ as the minimal effort required to reconfigure $\mu$'s mass distribution into that of $\nu$.

\begin{figure}
\begin{center}
\includegraphics[trim=0mm 123mm 0mm 0mm, clip, scale=0.8]{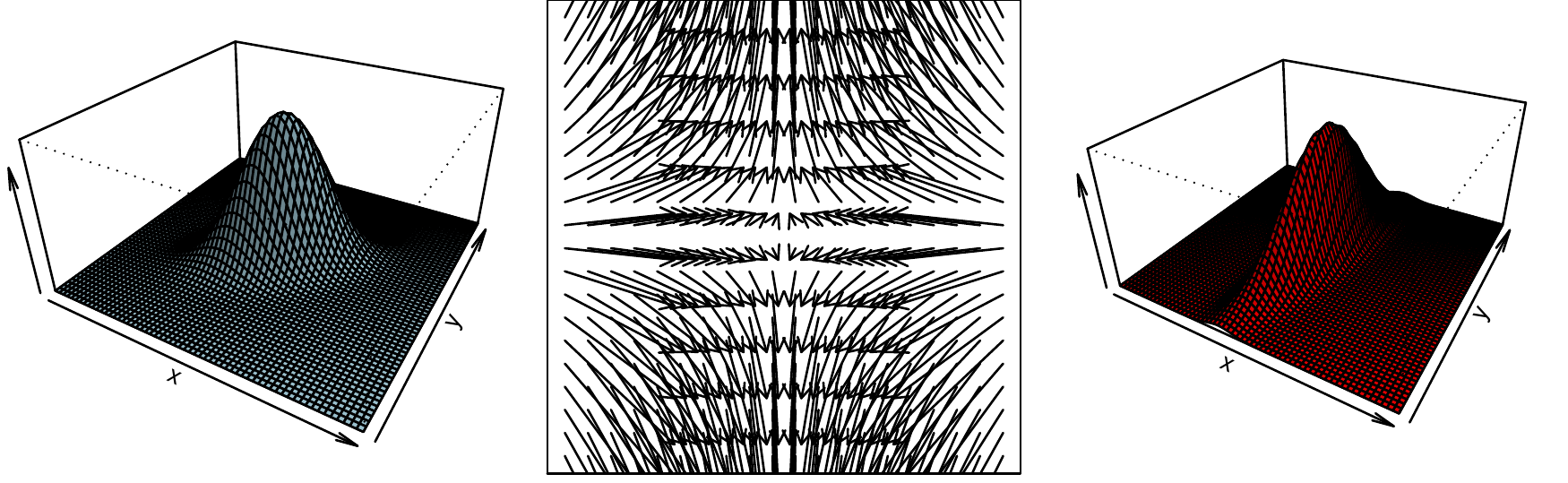}
\includegraphics[trim=0mm 121mm 0mm 0mm, clip, scale=0.8]{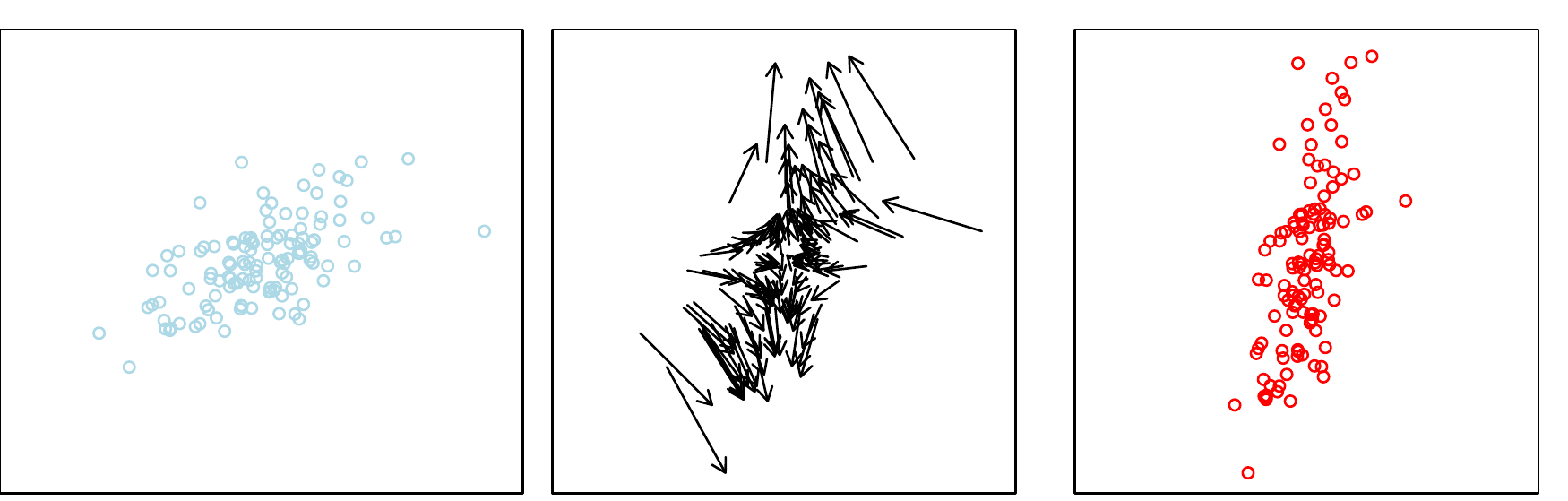}
\end{center}
\caption{Illustration of the ``analytic" and ``probabilistic" definitions. The top row of plots shows the densities of two Gaussian probability measures $\mu$ (on the left, in blue) and $\nu$  (on the right, in red), and the optimal deterministic map $T$ (in the middle) that deforms $\mu$ into $\nu$, i.e., $T\#\mu=\nu$. The map is plotted in the form of the vector field $T(x)-x$, where each arrow indicates the source and destination of the mass being transported. Reversing the direction of the arrows would produce the inverse map, optimally deforming the measure $\nu$ to obtain $\mu$. The bottom row features two independent random samples $X_1,\ldots,X_N\stackrel{\mathrm{i.i.d.}}{\sim}\mu$  (on the left, in blue) and $Y_1,\ldots,Y_N\stackrel{\mathrm{i.i.d.}}{\sim}\nu$ (on the right, in red), for $N=120$. The sample $\{X_i\}_{i=1}^{N}$ was constructed by sampling $\mu$ directly. The sample $\{Y_i\}_{i=1}^{N}$ was constructed by applying the optimal map $T$ to the sample $\{X_i\}_{i=1}^{N}$, i.e. $Y_i=T(X_i)$. The plot in the middle illustrates how the sample $\{X_i\}_{i=1}^{N}$ is re-arranged in order to produce the sample $\{Y_i\}_{i=1}^{N}$, by plotting the vectors $T(X_i)-X_i$. The optimality of $T$ can be understood in terms of minimising the average squared length of these arrows.  In all plots, the $x$ and $y$ axes range from $-3$ to $3$.}
\label{fig:illustrationTransport}
\end{figure}

Definition \ref{analyst_definition} underlines that the feasible set $\Gamma$ is convex and that the objective function is (up to the power $1/p$) linear in $\gamma$. Optimal $\gamma$'s can thus be expected to be extremal, that is, relatively \emph{sparse}.  Examples of such sparse couplings are \emph{deterministic} ones, i.e., couplings supported on the graph of some deterministic function $T:\X\to\X$, rather than on $\X\times \X$, so that they can be realised as
\[
\gamma(A\times B)=\mu(A\cap T^{-1}(B)).
\]
Such a coupling reassigns \emph{all} of $\mu$'s mass at a given location to a \emph{unique} destination. When the vector $(X,Y)$ is distributed according to such a $\gamma$, its two coordinates are completely dependent: $Y=T(X)$ for the deterministic function $T:\X\to\X$.  Such $T$ is called an \emph{optimal transport map} and must satisfy $\nu(B)=\mu(T^{-1}(B))$ for all $B\subseteq\X$ if $\gamma$ is to be in $\Gamma$, i.e., $T$ \emph{pushes $\mu$ forward to $\nu$} (denoted by $T\#\mu=\nu$). Figure~\ref{fig:illustrationTransport} illustrates these definitions. 

As it turns out, under sufficient regularity, it is precisely such deterministic couplings that are optimal.  When $\X=\R^d$ is finite-dimensional and $\mu$ is absolutely continuous with respect to Lebesgue measure, the infimum (if finite) is attained (uniquely if $p>1$) by such a deterministic coupling.  In this case we denote the map $T$ inducing the coupling by $\topt XY$ or $\topt\mu\nu$.  In the next paragraph we briefly sketch the arguments leading to this result.   As the problem is analytical in nature, characterising the solutions requires some tools from mathematical analysis.  We have attempted to avoid technicalities to the extent possible, but with optimal transport ``the devil is in the details", as the problem is qualitatively different depending on whether the random variables are discrete or continuous.  The less mathematically-inclined reader can skip to the paragraph containing Equation~\ref{eq:optimalMap}, simply retaining the loose statement that in the quadratic case $p=2$, optimal maps are characterised as gradients of convex functions.  Our presentation mainly follows \cite{villani2003topics}; more references are given at the end of this section.

\textbf{Uniqueness and characterisation.}  Like any convex optimisation problem, the Monge--Kantorovich problem admits a dual, consideration of which leads to a \emph{characterisation} of optimal maps. The dual problem can be seen to be
\[
\sup_{\phi,\psi}\Big\{ \E \phi(X) + \E \psi(Y)\Big\},
\qquad \textrm{subject to }\quad \phi(x) + \psi(y) \le \|x-y\|^p
\]
for integrable $\phi$ and $\psi$.  The inequality $\E \phi(X) + \E\psi(Y)\le \E\|X-Y\|^p$ implies \emph{weak duality}, in that the above supremum is no larger than the infimum in Definition \ref{prob_definition}.  But under mild conditions one has, in fact, \emph{strong duality}, and there exist a pair $(\phi,\psi)$ and a joint coupling $\gamma$ such that $\E \phi(X) + \E\psi(Y)=\E_\gamma\|X-Y\|^p$.  Furthermore, a version of \emph{complementary slackness} holds between the two optimal solutions, in such a way that one provides a lot of information on the other.  This is best demonstrated in the quadratic case $p=2$, by virtue of the factorisation $\|x-y\|^2=\|x\|^2 + \|y\|^2 - 2\innprod xy$.  Algebraic manipulations then allow the dual to be recast as
\[
\inf_{\varphi,\Psi}\Big\{ \E \varphi(X) + \E \Psi(Y)\Big\},
\qquad \textrm{subject to }\quad \varphi(x) + \Psi(y) \ge \innprod xy.
\]
A simple yet consequential observation is that for a given $\varphi$, the best candidate for $\Psi$ is the \emph{Legendre transform} of $\varphi$,
\[
\varphi^*(y)
=\sup_{x\in\X}\{ \innprod xy - \varphi(x)\},
\]
the smallest function satisfying $\varphi^*(y)+\varphi(x)\ge \innprod xy$.  Iterating this idea amounts to replacing $\varphi$ by $\varphi^{**}=(\varphi^*)^*$, which is larger than $\varphi$ yet still obeys the constraint $\varphi^{**}(x)+\varphi^*(y)\ge \innprod xy$.  The choice $\Psi=\varphi^*$ makes the dual \emph{unconstrained}, and $\varphi$ is optimal if and only if $\varphi(x)+\varphi^*(y)=\innprod xy$ with probability one with respect to $X$ and $Y$.  Going back to the primal problem, we see that once an optimal $\varphi$ is found, a joint distribution will be optimal if and only if it assigns unit probability to the event $\varphi(X)+\varphi^*(Y)=\innprod XY$.  Furthermore, $\varphi$ itself may be assumed to be the Legendre transform of $\varphi^*$, namely $\varphi=\varphi^{**}$.

At this stage one can invoke the rich theory of convex analysis.  Legendre transforms are always convex, and the equality $\varphi(x)+\varphi^*(y)=\innprod xy$ holds if and only if $y$ is a \emph{subgradient} of $\varphi$ at $x$.  If $\varphi$ has a unique subgradient $y$ at $x$, then $y=\nabla\varphi(x)$ is the gradient of $\varphi$ and is determined uniquely.  The regularity of convex functions implies that this is the case for all $x$ up to a set of Lebesgue measure 0.  Thus, if $X$ has a density, then the optimal map $T$ is characterised as the unique gradient of a convex function that pushes $X$ forward to $Y$.  On the other hand, if $X$ is discrete, then it might be concentrated on the small set where $\varphi$ is not differentiable, in which case the optimal coupling will not be induced from a map.

Similar arguemnts apply for other values of $p>1$.  For a given $\phi$, the best candidate for $\psi$ is the \emph{$c$-transform}\footnote{Here the cost of transferring a unit of mass from $x$ to $y$ is $c(x,y)=\|x-y\|^p$, but these ideas are valid for more general cost functions $c$, hence the name.} of $\phi$,
\[
\phi^c(y)
=\inf_{x\in\X}\{\|x-y\|^p - \phi(x)\},
\]
which again leads to an unconstrained dual problem $\sup_\phi \E\phi(X)+\phi^c(Y)$.  A function $\phi$ is optimal if and only if $\phi(x)+\phi^c(y)=\|x-y\|^p$ with probability one, and $\phi$ itself can be assumed a $c$-transform.  In analogy with the quadratic case, the equality $\phi(x)+\phi^c(y)=\|x-y\|^p$ entails a relation between $y$ and the gradient of $\phi$, and $c$-transforms enjoy differentiability properties similar to those of convex functions.

In summary, when $X$ has a density, optimal maps $\topt XY$ are precisely functions of the form
\begin{equation}\label{eq:optimalMap}
\topt XY(x)
=\begin{cases}
\nabla \varphi(x) \textrm{ for some convex }\varphi, & p=2,\\
x - \|\nabla \phi(x)\|^{1/(p-1) - 1}\nabla\phi(x) \textrm{ for some }c\textrm{-transform }\phi, & p\ne2.
\end{cases}
\end{equation}
This formula for general $p$ is also valid if $p=2$, with $\phi(x)=\|x\|^2/2-\varphi(x)$.  Importantly, this uniqueness and characterisation result holds for two classes of spaces $\X$ extending $\R^d$:  Riemmanian manifolds and separable Hilbert spaces.

\textbf{Regularity.}  The convex gradient characterisation gives rise to a rich regularity theory in the quadratic case.  When both $\mu$ and $\nu$ have densities $f$ and $g$, the convex potential $\varphi$ solves the Monge--Amp\`ere equation
\[
\mathrm{det}\nabla^2\varphi(x)
=\frac{f(x)}{g(\nabla\varphi(x))}.
\]
The regularity theory of Monge--Amp\`ere equations allows one to deduce smoothness of the optimal map $T=\nabla\varphi$.  Roughly speaking, if $X$ and $Y$ have convex supports and positive, bounded densities with derivatives up to order $k\ge0$, then the optimal map $\topt\mu\nu$ has continuous derivatives up to order $k+1$.

\textbf{Explicit solutions.}  Apart from the characterisation of optimal maps $T$ as gradients of convex functions (when $p=2$) or $c$-transforms, typically neither $T$ nor the Wasserstein distance $W_p$ admit closed-form expressions.  There are two special yet important cases where one does have explicit formulae.  When $d=1$, denoting $F_X$ and $F_X^{-1}(q)=\inf\{x:F_X(x)\ge q\}$, $q\in(0,1)$, the distribution and quantile functions of $X$, we have
\[
W_p(X,Y)
=\|F_X^{-1} - F_Y^{-1}\|_p
=\left(\ownint 01{|F_X^{-1}(\alpha) - F_Y^{-1}(\alpha)|^p}\alpha\right)^{1/p},
\qquad \topt XY=F_Y^{-1} \circ F_X,
\]
where $\topt XY$ is optimal if $X$ is a continuous random variable.  This allows the quantile function $F^{-1}_Y$ of any random variable $Y$ to be interpreted as the optimal map from a uniform random variable to $Y$ (also see Section \ref{sec:MoreReferences} for an interesting interpretation/extension of this fact).  In the special case $p=1$ there is an alternative, often more convenient, formula:
\[
W_1(X,Y)
=\ownint {\R}{}{|F_X(t) - F_Y(t)|}t.
\]
The function $\topt XY=F_Y^{-1}\circ F_X$ is still optimal, but might not be unique.  One can also bound $W_p$ in terms of the distribution functions:
\[
W^p_p(X,Y)
\le p2^{p-1}\ownint {\R}{}{|t|^{p-1}|F_X(t) - F_Y(t)|}t.
\]
The other case where closed-form formulae are available is when $X$ and $Y$ are Gaussian.  If $X\sim N(m_1,\Sigma_1)$ and $Y\sim N(m_2,\Sigma_2)$, then
\begin{equation}
\begin{aligned}\label{eq:gaussTransport}
W_2^2(X,Y)
&= \|m_1 - m_2\|^2  + \mathrm{tr}[\Sigma_1 + \Sigma_2 - 2(\Sigma_1^{1/2}\Sigma_2\Sigma_1^{1/2})^{1/2}],\\
\topt XY(x)
&=m_2 + \Sigma_1^{-1/2}[\Sigma_1^{1/2}\Sigma_2\Sigma_1^{1/2}]^{1/2}\Sigma_1^{-1/2}(x - m_1),
\end{aligned}
\end{equation}
where $\topt XY$ is defined if $\Sigma_1$ is injective (more generally, if its kernel is included in that of $\Sigma_2$).  These formulae are valid in infinite dimensions too, in which case $\topt XY$ may be unbounded, and only defined on an affine subspace of $\X$.  Furthermore, this result holds in location-scale families that are not necessarily Gaussian.

\subsection{Bibliographic Notes}
In addition to the survey \cite{rachev1985monge}, there are a number of books dedicated to optimal transport:
\cite{rachev1998mass}, \cite{villani2003topics,villani2008optimal}, \cite{ambrosio2013user}, \cite{santambrogio2015optimal},  and the forthcoming \cite{panaretos2018invitation}, leaning to the statistical side of the subject.  The reader interested in the extensive bibliography may consult in particular the first, second and fourth of these references.  For space considerations, we only give a very brief historical overview and a summary list of references.

The origin of the optimal transport problem is the monograph by \cite{monge1781memoire}, in which he posed the question for the particular case $\X=\R^3$ and $p=1$;  see also \cite{appell1887memoire} for an early reference.  The probabilistic formulation of \cite{kantorovich1942translocation} was a major breakthrough, and one of the catalysers that led Kantorovich to develop linear programming, for which he was awarded the Nobel prize in 1975 (jointly with T.~C. Koopman, who independently arrived at similar results after Kantorovich).

Duality results have a rich history dating back at least to \cite{kantorovich1958space}.  Very general results (for all Borel cost functions) in this context can be found in \cite{beiglbock2011duality}.  See also \cite{kellerer1984duality}, who explores duality in a \emph{multimarginal} formulation involving more than two measures (see also Section~\ref{sec:statWass}).

The one-dimensional case is intrinsically related to the Fr\'echet--Hoeffding bounds \citep{hoffding1940masstabinvariante,frechet1951tableaux}.  See \cite{bass1955compatibilite} and \cite{dallaglio1956sugli} for early references, and \cite{cuesta1993optimal} for detailed discussion when $p=2$.  The bound for $W_p$ in terms of distribution functions is due to \cite{ebralidze1971inequalities}, and can be found in generalised form in \citet[Section 7.4]{bobkov2014one}.  There are analogous results for measures on spaces with simple structure;  see \cite{delon2010fast} for the unit circle and \cite{kloeckner2015geometric} for ultrametric spaces.

For the Gaussian case, see \cite{olkin1982distance} or \cite{givens1984class} in finite dimensions, and \cite{gelbrich1990formula} and \cite{cuesta1996lower} for an infinite-dimensional extension.

The convex gradient characterisation in the quadratic case was discovered independently by a number of authors: \cite{knott1984optimal}, \cite{cuesta1989notes}, \cite{ruschendorf1990characterization} and \cite{brenier1991polar}.  For other values of the exponent $p$ (and more general cost functions), see \cite{gangbo1996geometry}.  The Riemannian version is due to \cite{mccann2001polar}, and \citet[Section~6.2.2]{ambrosio2008gradient} treat the infinite-dimensional case.

The regularity result was discovered by \cite{caffarelli1992regularity};  see \cite{figalli2017monge} for an accessible exposition.  There are other (e.g., Sobolev) types of regularity results, as explained in \citet[pages 332--336]{villani2008optimal} or \citet[Section 1.7.6]{santambrogio2015optimal}.

\section{Optimal Transport as a Technical Tool}\label{sec:tool}
This section reviews some of the features of Wasserstein metrics that make them useful as a technical tool for deriving large sample theory results in statistics.  To facilitate the presentation, we first state some simple facts that play a role in the development.  Let $X$ and $Y$ be random vectors taking values in $\X=\mathbb R^d$;  we maintain the notation $(\X,\|\cdot\|)$ to stress that the properties are valid in infinite dimensions as well.
\begin{itemize}
\item For any real number $a$, $W_p(aX,aY)=|a|W_p(X,Y)$.
\item For any fixed vector $x\in \X$, $W_p(X+x,Y+x)=W_p(X,Y)$.
\item For any fixed $x\in\X$, we have $W_2^2(X+x,Y)=\|x+\E(X)-\E(Y)\|^2+W_2^2(X,Y)$.
\item For product measures and when $p=2$, we have $W_2^2(\otimes_{i=1}^n \mu_i,\otimes_{i=1}^n\nu_i)=\sum_{i=1}^n W_2^2(\mu_i,\nu_i)$ in the analytic notation.
\end{itemize}
The proofs of the first three statements rely on the equivalence between the classes of the corresponding couplings.  For example, $U=(X+x,Y+x)$ is a coupling of $X+x$ and $Y+y$ if and only if $U-(x,x)$ is a coupling of $(X,Y)$.  For the last property, observe that the map $x\mapsto [\topt{\mu_1}{\nu_1}(x),\dots,\topt{\mu_1}{\nu_1}(x)]$ is a gradient of a convex function and pushes forward $\otimes\mu_i$ to $\otimes \nu_i$.

\subsection{Deviations from Gaussianity}
If $\{X_i\}_{i\geq 1}$ are independent and identically distributed random variables with mean zero and finite variance, then the central limit theorem asserts that the suitably rescaled averages $S_n= n^{1/2}\overline X_n$ converge in distribution to a normal random variable $Z$ with the same variance.  Since $\E S_n^2=\E Z^2$, the convergence also holds in 2-Wasserstein distance.  This property makes the 2-Wasserstein distance convenient for handling deviations from Gaussianity.  The arguments generally involve the \emph{subadditivity} of the Wasserstein distance with respect to convolutions, a property that can be established using the infimum-over-couplings definition of the Wasserstein distances.  For example, assuming $\E X_i=0$,
\begin{equation}
\label{eq:subad}
W_2^2\left(\sum_{i=1}^n a_iX_i,Z\right)
\le \sum_{i=1}^n a_i^2 W_2^2(X_i,Z)
,\qquad Z\sim N(0,1)
,\qquad 
\sum_{i=1}^n a_i^2=1. 
\end{equation}
To see this, let $Z_i\sim N(0,1)$ be independent and consider optimal couplings on $\R^2$ such that $\E |a_iX_i - a_iZ_i|=W_2^2(a_iX_i,a_iZ_i)$.  Take the product $\pi$ of all these couplings (a joint distribution on $\R^{2n}$).  Then under $\pi$, $\sum a_iZ_i$ is standard normal and
\[
W_2^2\left(\sum_{i=1}^n a_iX_i,Z\right)
\le \E_\pi \left|\sum_{i=1}^n a_iX_i - \sum_{i=1}^n a_iZ_i\right|^2
=\sum_{i=1}^n \E \left|a_iX_i - a_iZ_i\right|^2
=\sum_{i=1}^n W_2^2(a_iX_i,a_iZ),
\]
from which Equation \ref{eq:subad} follows.  \cite{mallows1972note} used this property in order to derive necessary and sufficient conditions for a triangular array to be \emph{jointly asymptotically normal}.  Recall that $X_n=(X_{n1},\dots,X_{nd})$ converge in distribution to a standard multivariate $N(0,I_d)$ if and only if $a^tX_n\to Z$ for all $a\in \R^d$, $\|a\|=1$.  Now let $X_{nj}$ ($j\le n<\infty$) be a triangular array.  In analogy with a fixed dimension, we say that $(X_{nj})$ is jointly asymptotically normal if $a_n^tX_n\to Z$ for any sequence of vectors $a_n\in \R^n$, $\|a_n\|=1$.  This requires $X_{nj}$ to converge to $Z$ uniformly in $j$, i.e., $X_{nm_n}\to Z$ for any sequence of coordinates $m_n\le n$.  This condition is not sufficient, however.  \cite{mallows1972note} observed that metrics inducing convergence in distribution are not subadditive, and this is remedied by the Wasserstein distance.  If $\E X_{nj}^2\to1$ uniformly in $j$, in addition to the uniform convergence in distribution, then $W_2^2(X_{nj},Z)\to0$ and as a consequence of Equation~\ref{eq:subad}, $W_2^2(a_n^tX_n,Z)\to0$, and the array is jointly asymptotically normal. The length of the $n$-th row of the array can be arbitrary, as long as it diverges to infinity with $n$.

When the $X_i$'s in Equation~\ref{eq:subad} have the same distribution as $X$ and $a_i=1/\sqrt n$, the inequality gives a bound that is uniform in $n$.  \cite{bickel1981some} use this result in their study of the asymptotics of the bootstrap.  For instance, denote by $F_n$ the empiricial distribution function corresponding to a sample $X_1,\dots,X_n$ and the sample mean by $\mu_n=\overline X$.   Let $X_1^*,\dots,X_m^*$ be a bootstrapped sample from $F_n$ with sample average $\mu_m^*$.  Then as $n,m\to\infty$, the conditional (upon $(X_i)$) distribution of $\sqrt m(\mu_m^* - \mu_n)$ converges to $N(0,\mathrm{var}(X_1))$, which is the same asymptotic distribution of $\mu_n$.

Another additive property, shown in a similar way to Equation \ref{eq:subad}, is
\[
W_p\left(\sum_{i=1}^n U_i,\sum_{i=1}^n V_i\right)
\le \sum_{i=1}^n W_p(U_i,V_i),
\]
for independent $(U_i)$ and $(V_i)$.   A particular case is that $W_p(X+Y,X)\le W_p(Y,0)=[\E\|Y\|^p]^{1/p}$, and taking $Y$ to be Gaussian with small variance allows to approximate in $W_p$ any probability law with a smooth one to arbitrary precision.  In other words, smooth measures are dense in $W_p$, just as they are dense with respect to convergence in distribution.  Discrete measures are also dense;  see Subsection~\ref{sec:empiricalWass}.

Actually, the subadditivity properties can be used in order to prove the central limit theorem.  \cite{tanaka1973inequality} does so by noticing that equality in (\ref{eq:subad}) holds only for Gaussian distributions.  \cite{johnson2005central} obtain rates of convergence for the central limit theorem, and more generally, for convergence to stable laws.  Berry--Esseen-type bounds for the Wasserstein distance can be found in \cite{rio2009upper}.  For random elements in Banach spaces, see \cite{rachev1994rate}.

\subsection{Equilibrium, Concentration, and Poisson Approximations}

A different class of settings where Wasserstein distances are used is in the study of convergence of Markov chains to their equilibrium distribution and dates back to \cite{dobrushin1970prescribing}.  The idea is to show a sort of contraction property of the transition kernel with respect to the Wasserstein distance.  Let $P$ be the transition matrix.  In studying convergence of the Kac random walk on the orthogonal group $\mathrm{SO}(n)$, \cite{oliveira2009convergence} showed that
\[
W_{D,2}(\mu P,\nu P)
\le \xi W_{D,2}(\mu,\nu)
\]
for some $\xi<1$, where $D$ is a distance between matrices, leading to exponential convergence to equilibrium.  A result of similar spirit is derived by \cite{eberle2014error} for the transition kernel of the Metropolis adjusted Langevin algorithm, a Markov chain Monte Carlo method.  The constant $\xi$ above is related to the \emph{Wasserstein spectral gap} of the transition kernel.  \cite{hairer2014spectral} explore its behaviour  in infinite-dimensional state spaces, when taking finite-dimensional projections of $P$.  They show that for the preconditioned Crank--Nicolson algorithm, $\xi$ remains stable, whereas for the random walk Metropolis algorithm, $\xi$ may converge to one.  \cite{rudolf2018perturbation} employ Wasserstein distances to bound the difference between the behaviour of some ``nicely behaved" Markov chain and a perturbed version thereof, obtained from a modification in the transition kernel.

Wasserstein distances also appear in concentration of measure, in the form of \emph{transportation inequalities} \cite[Chapter 6]{ledoux2005concentration}.  A measure $\mu_0$ satisfies such an inequality if for any other measure $\nu$,
\[
W_1(\mu_0,\nu)
\le C\sqrt{H(\mu_0,\nu)},
\qquad H(\mu,\nu)
=\ownint{}{}{\log\frac{\rm d\mu}{\rm d\nu}}\mu.
\]
If this holds, and $\mu(A)\ge1/2$, then
\[
\P(X \notin A_r)
\le e^{-r^2/C'},
\qquad A_r=\{x:\|x-A\|\le r\}. 
\]
Furthermore, the representation of $W_1$ as the supremum over Lipschitz functions (see the next subsection) yields concentration inequalities for $f(X) - \E f(X)$ with $f$ Lipschitz.

In a different context, \cite{barbour1992stein} use Wasserstein metrics to quantify the error in approximating a point process $\Xi$ by a Poisson point process $P$ with the same mean measure $\lambda$.  Suppose for simplicity that the sample space is $[0,1]$ and for two (not necessarily probability) measures $\tilde\mu,\tilde\nu$ with total masses $A$ and $B$, define the probabilities $\mu=\tilde\mu/A$, $\nu=\tilde\nu/B$ and $d(\tilde\mu,\tilde\nu)=W_1(\mu,\nu)$ if $A=B$ and 1 (the maximal value) if $A\ne B$.  The processes $\Xi$ and $P$ can then be viewed as random elements in the metric space $\X$ of measures with the distance $d$, and their laws can be compared using the ``upper degree" Wasserstein space $W_1$ on $(\X,d)$.  See also \cite{schuhmacher2009stein} for an extension where $d$ is replaced by a Wasserstein distance of different order $W_p$.

\subsection{Relation to Other Metrics}
We conclude this section by reviewing some useful relations between $W_p$ and other probability metrics.  We firstly relate $W_p$ to $W_q$ by two simple results from \citet[Chapter 7]{villani2003topics}, and then describe bounds (mostly borrowed from \cite{gibbs2002choosing}) pertaining to $W_1$ and the Prokhorov, total variation and bounded Lipschitz distances.  For notational simplicity we state the bounds in the Euclidean setting, but they hold on any complete separable metric space $(\X,\rho)$.  For random variables $X$ and $Y$ on $\X$ let $\Omega$ be the union of their ranges and set
\[
D
=\sup_{x,y\in \Omega}
\|x-y\|,
\qquad d_{\min}
=\inf_{x\ne y\in\Omega}
\|x-y\|.
\]
In the analytic version $\Omega=\mathrm{supp}(\mu)\cup \mathrm{supp}(\nu)$, where $X\sim \mu$ and $Y\sim \nu$.  If $X$ and $Y$ are bounded, then $D$ is finite;  if $X$ and $Y$ are (finitely) discrete, then $d_{\min}>0$.

\begin{itemize}
\item If $p\le q$, then $W_p\le W_q$, by Jensen's inequality.
\item On the other hand, $W_q^q\le W_p^pD^{q-p}$.
\item Duality arguments yield the particularly useful \emph{Kantorovich--Rubinstein} \citep{kantorovich1958space} representation for $W_1$ as
\[
W_1(X,Y)
=\sup_{\|f\|_{\mathrm{Lip}}\le 1} |\E f(X) - \E f(Y)|,
\qquad \|f\|_{\mathrm{Lip}}
=\sup_{x\ne y}\frac{|f(x) - f(y)|}{\|x - y\|},
\]
valid on any separable metric space \cite[Section 11.8]{dudley2002real}.
\item This shows that $W_1$ is larger than the \emph{Bounded Lipschitz} (BL) metric
\[
W_1(X,Y)
\ge
\mathrm{BL}(X,Y)
=\sup_{\|f\|_\infty+\|f\|_{\mathrm{Lip}}\le 1} |\E f(X) - \E f(Y)|
\]
that metrises convergence in distribution \cite[Theorem 11.3.3]{dudley2002real}.
\item Let $P$ denote the Prokhorov distance.  Then $P^2(X,Y)\le W_1(X,Y)\le (D+1)P(X,Y)$.
\item For the class of random variables supported on a fixed bounded subset $K\subseteq \X$, $\mathrm{BL}$ and $W_1$ are equivalent up to constant, and all metrics $W_p$ are topologically equivalent.
\item The Wasserstein distances $W_p$ can be bounded by a version of total variation $\mathrm{TV}$ \cite[Theorem 6.15]{villani2008optimal}.  A weaker but more explicit bound for $p=1$ is $W_1(X,Y)\le D\times \mathrm{TV}(X,Y)$.
\item For discrete random variables, there is an opposite bound $\mathrm{TV}\le W_1/d_{\min}$.
\item The total variation between convolutions with a sufficiently smooth measure is bounded above by $W_1$ \citep[Proposition~4]{mariucci2017wasserstein}.
\item The \emph{Toscani} (or \emph{Toscani--Fourier}) distance is also bounded above by $W_1$ \citep[Proposition~2]{mariucci2017wasserstein}.
\end{itemize}
Beyond bounded random variables, $W_p$, $W_q$, $\mathrm{BL}$ and $\mathrm{TV}$ induce different topologies, so that one cannot bound, for example, $W_1$ in terms of $\mathrm{BL}$ in the unbounded case.  On a more theoretical note, we mention that the Kantorovich--Rubinstein formula yields an embedding of any Polish space $(\X,\rho)$ in the Banach space of finite signed measures on $\X$.

\section{Optimal Transport as a Tool for Inference}\label{sec:inference}
As a measure of distance between probability laws, the Wasserstein distance can be used for carrying out of goodness-of-fit tests, and indeed this has been its main use as a tool for statistical inference.  In the simplest \emph{one-sample} setup, we are given a sample $X_1,\dots,X_n$ with unknown law $\mu$ and wish to test whether $\mu$ equals some known fixed law $\mu_0$ (e.g., standard normal or uniform).  The \emph{empirical measure} $\mu_n$ associated with the sample $(X_1,\dots,X_n)$ is the (random) discrete measure that assigns mass $1/n$ to each observation $X_i$.  In this sense, the strong law of large numbers holds in Wasserstein space:  with probability one, $W_p(\mu_n,\mu)\to0$ as $n\to\infty$ if and only if $\E \|X\|^p<\infty$.  It is consequently appealing to use $W_p(\mu_n,\mu_0)$ as a test statistic.  In the \emph{two-sample} setup, one independently observes a sample $Y_1,\dots,Y_m\sim \nu$ with corresponding empirical measure $\nu_m$, and $W_p(\mu_n,\nu_m)$ is a sensible test statistic for the null hypothesis $\mu=\nu$.

\subsection{Univariate Measures}
We shall identify measures $\mu$ on the real line ($\X=\R$), with their distribution function $F$; the \emph{empirical distribution function} corresponding to $\mu_n$ is $F_n(t)=n^{-1}\sum_{i=1}^n\mathbf{1}\{X_i\le t\}$.  Thus $X_i\sim F$, $Y_j\sim G$ and we slightly abuse notation by writing $W_p(F,G)$ for $W_p(\mu,\nu)$.

\cite{munk1998nonparametric} derive the asymptotic distribution of $W_2(F_n,F_0)$ (and trimmed versions thereof).  The main tool for the derivation is a Brownian bridge representation for the quantile process $q_n=\sqrt n(F_n^{-1} - F^{-1})$ that holds under suitable assumptions on $F$.  There are four types of limiting results, depending on the combination null/alternative and one/two-sample.  Roughly speaking, the limits are of order $\sqrt n$ and normal under the alternative, and of order $n$ and not normal under the null.  The two-sample asymptotics entail that $m/n$ converges to a finite positive constant.  In symbols:
\begin{equation}\label{eq:fourlimits}
\begin{aligned}
\sqrt n(W_2^2(F_n,F_0) - W_2^2(F,F_0))&\to \textrm{normal} \quad (F\ne F_0),\\
nW_2^2(F_n,F_0)&\to 
\textrm{something} \quad (F=F_0),\\
\sqrt \frac {mn}{m+n}(W_2^2(F_n,G_m) - W_2^2(F,G))&\to \textrm{normal} \quad (F\ne G),\\
 \frac {mn}{m+n} W_2^2(F_n,G_m) &\to \textrm{something} \quad (F=G).
\end{aligned}
\end{equation}
Similar results were obtained independently in \cite{delBarrio2000contributions}, where one can also find a nice survey of other goodness-of-fit tests.

If one instead wants to test whether $F$ belongs to a parametric family $\mathcal F$ of distributions, then the test statistic is the infimum of the Wasserstein distance between the empirical measure and members of $\mathcal F$.  For example, in order to test the fit to some normal distribution, \cite{delBarrio1999tests} find the asymptotic distribution of the test statistic
\[
R_n = \frac{\inf_{\mu,\sigma^2}W^2_2(F_n,N(\mu,\sigma^2))}{S_n^2},
\qquad S_n^2 = \frac1n\sum_{i=1}^n (X_i - \overline X)^2,
\]
an infinite sum of rescaled and centred $\chi^2$ random variables (under the null hypothesis).  Using a weighted version of the Wasserstein distance, \cite{deWet2002goodness} constructs a test for location or scale families.  Here the null hypothesis is that $F=F_0(\cdot - \theta)$ or $F=F_0(\cdot/\theta)$ for some known distribution $F_0$ and $F$ and unknown $\theta\in \R$ (or $(0,\infty)$).  In a more general setup, \cite{freitag2005hadamard} consider the case of a ``structural relationship" between $F$ and $F_0$ in the form
\[
F^{-1}(t)
=
\phi_1(F_0^{-1}(\phi_2(t,\theta)),\theta),
\]
for some (known) functions $\phi_1,\phi_2:\R\times \Theta\to\R$ and parameters $\theta\in\Theta$.  This setup includes the location-scale model when $\phi_2(t,\theta)=t$ and $\phi_1(t,\theta_1,\theta_2)=(t-\theta_1)/\theta_2$, and the \emph{Lehmann alternatives} model when $\phi_2(t,\theta)=1-(1-t)^\theta$ and $\phi_1(t,\theta)=t$.  Motivated by population bioequivalence problems, \cite{freitag2007nonparametric} treat the dependent two-sample case, where one observes a sample $(X_i,Y_i)_{i=1}^n$ and wishes to compare the Wasserstein distance between the marginals.

Some of the required regularity is apparent from the following observation.  The empirical process $\sqrt n(F_n - F)$ converges to $\mathbb B\circ F$, where $\mathbb B$ is a Brownian bridge on $[0,1]$, without assumptions on $F$ (this result is known as \emph{Donsker's theorem}).  But the quantile process $q_n$ involves inversion, and the limiting distribution is $\mathbb B(t)/F'(F^{-1}(t))$, which requires assumptions on $F$.  See \cite{csorgo1993weighted} for a detailed study of the quantile process and asymptotics of functionals thereof.  In the context of Wasserstein distance, \cite{delBarrio2005asymptotics} study the limiting behaviour of the norm $\|q_n\|_{2,w}^2=\ownint01{q_n^2(t)w(t)}t$, for an integrable weight function $w$ on $(0,1)$.  The covariance function of the process $\mathbb B/F'\circ F^{-1}$ is
\[
\eta(s,t)
=\frac{\min(s,t) - st}{F'(F^{-1}(t) F'(F^{-1}(s))},
\qquad s,t\in (0,1),
\]
and the limits are qualitatively different depending on whether the integrals $\ownint 01{\eta(t,t)w(t)}t$ and/or $\ownint 01{\ownint01{\eta^2(t,s)w(t)w(s)}t}s$ are finite or not.

\subsection{Multivariate Measures}
Results in the multivariate setup are more scarce.  One apparent reason for this is that the Wasserstein space of measures with multidimensional support is no longer embeddable in the function space $L_p(0,1)$ via quantile functions, and has positive curvature (see Section~\ref{sec:statWass}).  As perhaps can be expected, multivariate distributional results for the empirical $p$-Wasserstein distance are chiefly available when it admits a closed form;  that is, when $p=2$ and we consider Gaussian distributions.  Assume that $\mu=N(m_1,\Sigma_1)$.  Given a sample $X_1,\dots,X_n$ from $\mu$, let $\widehat{\mu}_n$ be the \emph{empirical Gaussian measure}
\[
\widehat\mu_n
=N(\widehat m,\widehat{\Sigma}),
\qquad \widehat m=\overline X=\frac 1n\sum_{i=1}^n X_i,
\quad \widehat \Sigma = \frac 1{n-1} \sum_{i=1}^n (X_i - \overline X)(X_i - \overline X)^t.
\]
The test statistic is now $W_2^2(\widehat{\mu}_n,\mu_0)$ for one sample and $W_2^2(\widehat{\mu}_n,\widehat{\nu}_m)$ for two samples, and the analogue of the four cases in Display \ref{eq:fourlimits} holds true.  The underlying idea is to combine the classical central limit theorem for $\widehat m$ and $\widehat\Sigma$ with a delta method, and \cite{rippl2016limit} establish the necessary differentiability of the squared Wasserstein distance in the Gaussian setup in order to apply the delta method.  Importantly, Gaussianity can be replaced with any location-scatter family of $d$-dimensional distribution functions
\[
\{F(x)
=F_0(m+\Sigma^{1/2} x)
:m\in \R^d; \Sigma\in\R^{d\times d} \textrm{ positive definite}
\},
\]
where $F_0$ is an arbitrary distribution function with finite nonsingular covariance matrix.

For sufficiently smooth measures $\mu,\nu$ (with moment conditions), \cite{delBarrio2017central} find the normal limit of 
\[
\sqrt n(W_2^2(\mu_n,\nu) - \E W_2^2(\mu_n,\nu)).
\]
They establish stability of the convex potential with respect to perturbations of the measures and invoke the Efron--Stein inequality.  Again in analogy with Display~\ref{eq:fourlimits}, the limiting distribution is degenerate at 0 if $\mu=\nu$.  This central limit theorem does not, however, yield a limit for $W_2^2(\mu_n,\nu) - W_2^2(\mu,\nu)$, since the speed at which $\E W_2^2(\mu_n,\mu)$ decays to zero (and consequently that of $\E W_2^2(\mu_n,\nu) - W_2^2(\mu,\nu)$) depends on $\mu$ in a rather delicate way, and can be arbitrarily slow (see Subsection~\ref{sec:empiricalWass}).

When $\mu$ and $\nu$ are finitely supported measures, they can be identified with vectors $r$ in the unit simplex, and the empirical vector $r_n$ obeys a central limit theorem.  \cite{sommerfeld2018inference} apply a delta method to obtain the limiting distributions of the Wasserstein distance.  The latter is only \emph{directionally Hadamard differentiable}, leading to a non-standard delta method with nonlinear derivative.   Correspondingly, the limiting distributions are not Gaussian, in general.  In analogy with Display~\ref{eq:fourlimits}, they show that $n^{1/2}(W_p(r_n,s) - W_p(r,s))$ has a distributional limit under the alternative, whereas under the null, the rate is $n^{1/(2p)}$ in agreement with results in Subsection~\ref{sec:empiricalWass}.  \cite{sommerfeld2018inference} highlight the implications of the non-standard delta method for the bootstrap, whose consistency requires subsampling.

These results extend to \emph{countably supported} measures, where one needs to impose an extra summability condition on $r$ in order to ensure convergence of $\sqrt n(r_n - r)$ to the Gaussian limit $\mathbb G$ \citep{tameling2017empirical}.  Both references also provide more explicit expressions for the limiting distribution when $\X$ has a metric structure of a tree.   \cite{bigot2017central} establish similar limits for a regularised version (see Section~\ref{sec:numerics}) of the Wasserstein distance.

Wasserstein distances have recently been proposed by \cite{bernton2017inference} for parameter inference in approximate Bayesian computation (also known as \emph{plug-and-play} methods).  The setup is that one observes data on $\X$ and wishes to estimate the underlying distribution $\mu$ belonging to a parametrised set of distributions $\{\mu_\theta\}_{\theta\in\R^N}$.  However, the densities of these measures are too complicated to evaluate/optimise a likelihood. Instead one can only simulate from them, and retain parameters that yield synthetic data resembling the observed data. A core issue here is how to contrast the true and simulated data, and  \cite{bernton2017inference} suggest using $W_p$ to carry out such comparisons.

A Wasserstein metric has also been employed to compare \emph{persistence diagrams}, a fundamental tool in topological data analysis (see \cite{wasserman2018topological} for a recent review) summarising the persistent homology properties of a dataset. See, for example,  \cite{mileyko2011probability}, who introduce a version of the Wasserstein distance on the space of persistence diagrams, endowing it with a metric structure that allows statistical inference.

\subsection{Bounds for the Expected Empirical Wasserstein Distance}\label{sec:empiricalWass}
As discussed in the previous subsections, the speed of convergence of the empirical measure $\mu_n$ to $\mu$ in Wasserstein distance $W_p$ is important for statistical inference. This topic has a history dating back to the seminal work of \cite{dudley1969speed}, and a very rich literature.  For space considerations, we will focus on the average value $\E W_p(\mu_n,\mu)$, but see the bibliographical notes for concentration inequalities and almost sure results.  Upper bounds for the one-sample version are also valid for the two-sample version, since $\E W_p(\mu_n,\nu_n)\le 2\E W_p(\mu_n,\mu)$ when $\nu_n$ is another empirical measure.  For brevity we write $W_p$ for $W_p(\mu_n,\mu)$ and inequalities such as $\E W_p\ge Cn^{-1/2}$ hold for given $p$, some $C=C(\mu)$ and for all $n$.  We also tacitly assume that $\mu\in \W_p$, i.e., it has a finite $p$-th moment, when writing $W_p$.

The behaviour of $\E W_p(\mu_n,\mu)$ is qualitatively different depending on whether the underlying dimension $d>2p$ or $d<2p$.  For discrete measures, $\E W_p$ is generally of the order $n^{-1/(2p)}$, independently of the dimension.  In high dimensions this is better than absolutely continuous measures, for which the rate is $n^{-1/d}$, but when $d=1$, some smooth measures attain the optimal rate $n^{-1/2}$, faster than $n^{-1/(2p)}$.  We first note that it is quite easy to see that $W_p\to0$ almost surely.  However, even for $p=1=d$ the decay of $\E W_p$ can be arbitrarily slow;  see \citet[Theorem 3.3]{bobkov2014one}.

Lower bounds are easier to obtain, and here are some examples:
\begin{itemize}
\item \emph{[fundamental $\sqrt n$ bound]} If $\mu$ is nondegenerate, then $\E W_p\ge Cn^{-1/2}$.
\item \emph{[separated support]} If $\mu(A)>0$, $\mu(B)>0$, $\mu(A\cup B)=1$ and
$dist(A,B)=\inf_{x\in A,y\in B}\|x-y\|>0$, then $\E W_p\ge C_pn^{-1/(2p)}$.  Any finitely discrete nondegenerate measure satisfies this condition, as well as most countably discrete ones.  This agrees with the rates of \cite{sommerfeld2018inference} above.
\item \emph{[curse of dimensionality]} If $\mu$ is absolutely continuous on $\R^d$, then $\E W_p\ge Cn^{-1/d}$.  (This result is void of content when $d\le 2$ in view of the $n^{-1/2}$ bound.)  More generally, $\mu$ only needs to have an absolutely continuous part (e.g.\ a mixture of a Gaussian with a discrete measure), and the bound holds when $\mu_n$ is replaced with \emph{any} measure supported on $n$ points.  Equivalently, it holds for the \emph{quantiser} of $\mu$, the $n$-point measure that is $W_p$-closest to $\mu$.
\end{itemize}
We briefly comment on how these bounds are obtained.  The $\sqrt n$ bound is a corollary of the central limit theorem on $f(X)$, where $X\sim \mu$ and $f$ is a suitable Lipschitz function.  If $\mu$ has separated support and $k\sim B(n,\mu(A))$ is the number of points in $\mu_n$ falling in $A$, then a mass of $|k/n - \mu(A)|$ must travel at least $dist(A,B)>0$ units of distance, yielding a lower bound on the Wasserstein distance.  One then invokes the central limit theorem for $k$.  For the curse of dimensionality, note that the number of balls of radius $\epsilon$ needed to cover the support of $\mu$ is proportional to $\epsilon^{-d}$.  If we take $\epsilon=Kn^{-1/d}$ with an appropriate $K>0$, then $n$ balls of radius $\epsilon$ centred at the points of the empirical measure miss mass $\tau$ from $\mu$, and this mass has to travel at least $\epsilon$, so $W_p^p\ge C'\tau n^{-p/d}$.

The last lower bound was derived by counting the number of balls needed in order to cover $\mu$, which turns out to be a determining quantity for the upper bounds, too.  To account for unbounded supports we need to allow covering only a (large) fraction of the mass.  Let
\[
N(\mu,\epsilon,\tau)
=\textrm{minimal number of }\epsilon\textrm{-balls whose union has }\mu\textrm{ measure at least }1-\tau.
\]
These \emph{covering numbers} increase as $\epsilon$ and $\tau$ approach zero, and are finite for all $\epsilon,\tau>0$.  To put the next upper bound in context, we remark that any compactly supported $\mu$ on $\R^d$ satisfies $N(\mu,\epsilon,0)\le K\epsilon^{-d}$.
\begin{itemize}
\item If for some $d>2p$, $N(\mu,\epsilon,\epsilon^{dp/(d-2p)})\le \epsilon^{-d}$, then $\E W_p\le C_pn^{-1/d}$.
\end{itemize}
This covering number condition is verified if $\mu$ has finite moment of order large enough \citep[Proposition 3.4]{dudley1969speed}.

The exact formulae on the real line lead to a characterisation of the measures attaining the optimal $n^{-1/2}$ rate:
\begin{itemize}
\item If $\mu\in \W_p(\R)$ has compact support, then $\E W_1\le Cn^{-1/2}$, and consequently $\E W_p\le C_pn^{-1/(2p)}$.
\item A necessary and sufficient condition for $\E W_1\le Cn^{-1/2}$ is that
\[
J_1(\mu)
=
J_1(F)
=
\ownint \R {}{\sqrt{F(t)(1-F(t))}}t
<
\infty.
\]
\item The same holds for $\E W_p$, with the integrand in $J_1$ replaced by $ [F(t)(1-F(t))]^{p/2}/[f(t)]^{p-1}$, where $f$ is the density of the absolutely continuous part of $\mu$.
\end{itemize}
Using the representation of $W_1$ as the integral of $|F_n - F|$, one sees that $J_1<\infty$ suffices for the $n^{-1/2}$ rate, since the integrand has variance $n^{-1}F(t)(1-F(t))$.  The condition $J_1<\infty$ is essentially a moment condition, as it implies $\E X^2<\infty$ and is a consequence of $\E X^{2+\delta}$ for some $\delta>0$.  But for $p>1$, $J_p<\infty$ entails some smoothness  of $\mu$.  In particular, the above lower bounds show that $\mu$ must be supported on an (possibly unbounded) interval, and the $J_p$ condition means that the density should not vanish too quickly in the interior of the support.

\subsubsection{Bibliographic Notes}
The lower bounds were adapted from \cite{dudley1969speed}, \cite{fournier2015rate} and \cite{weed2017sharp}.

The upper bound with the coverings dates back to \cite{dudley1969speed}, who showed it for $p=1$ and with the bounded Lipschitz metric.  The version given here can be found in \cite{weed2017sharp} and extends \cite{boissard2014mean}.  We emphasise that their results are not restricted to Euclidean spaces.  For Gaussian measures in a Banach space, \cite{boissard2014mean} relate $\E W_2$ to small ball probabilities.  \cite{weed2017sharp} also show that absolutely continuous measures that are ``almost" low-dimensional enjoy better rates for moderate values of $n$, until eventually giving in to the curse of dimensionality.

In the limiting case $d=2p$, there is an additional logarithmic term.  For $p=1$ the sufficiency of this term was noted by \citet[page 44]{dudley1969speed}, and the necessity follows from a classical result of \cite{ajtai1984optimal} for $\mu$ uniform on $[0,1]^2$.  For $p>1$ and $d=2p$, see for example \cite{fournier2015rate}.

That absolutely continuous measures are the ``bad" measures in high dimensions was already observed by \cite{dobric1995asymptotics} in an almost sure sense:  $n^{1/d}W_p$ has a positive limit if and only if $\mu$ has an absolutely continuous part.  There are results for more general cost functions than powers of Euclidean distance, see \cite{talagrand1994transportation} for $\mu$ uniform on $[0,1]^d$ and \cite{barthe2013combinatorial} for a careful study of the two-sample version $W_p(\mu_n,\nu_n)$.  \cite{fournier2015rate} also deal with the Euclidean case, with some emphasis on deviation bounds and the limit cases $d=2p$.

\cite{delBarrio1999central} showed that $J_1<\infty$ is necessary and sufficient for the empirical process $\sqrt n(F_n - F)$ to converge in distribution to $\mathbb{B}\circ F$, with $\mathbb B$ Brownian bridge.  A thorough treatment of the univariate case, including but not restricted to the $J_p$ condition, can be found in \cite{bobkov2014one}, using an order statistic representation for the Wasserstein distance.  One may also consult \cite{mason2016weighted} for the alternative approach of weighted Brownian bridge approximations.

The topic is one of intense study, and the references here are far from being exhaustive.  Let us also mention some extensions for dependent data: \cite{dede2009empirical}, \cite{cuny2017invariance}, \cite{dedecker2017behavior}.

\section{Optimal Transport as the Object of Inference}\label{sec:statWass}
The previous section described applications of Wasserstein distances for carrying out statistical tasks such as goodness-of-fit testing.  The topic of this section is a more recent trend, where one views the Wasserstein space as a sample space for statistical inference.  In this setup, one observes a sample $\mu_1,\dots,\mu_n$ from a random measure $\Lambda$ taking values in Wasserstein space $\W_p$ of measures with finite $p$-th moment, and seeks to infer some quantity pertaining to the law of $\Lambda$ using the observed data, typically in a nonparametric fashion. Such questions can be seen as part of \emph{next-generation functional data analysis}, borrowing the terminology of \citet[Section~6]{wang2016functional}.

\subsection{Fr\'echet Means of Random Measures}\label{subsec:frechetmeans}
Perhaps the most basic question here, as anywhere, is estimating a mean.  Clearly we could estimate the mean of $\Lambda$ by the average $n^{-1}(\mu_1+\dots+\mu_n)$, which is also a probability measure.  While this may often be a good estimator, in certain modern applications, such as imaging, it exhibits some unsatisfactory properties.  As a simple example, consider two Dirac measures at distinct points $x\ne y$.  Their average is the ``blurred" measure putting mass $1/2$ at $x$ and $y$.  In contrast, as we shall see below, the Wasserstein distance leads to an average that is a Dirac measure at the midpoint $(x+y)/2$.

We shall focus on the special case $p=2$, which is the most elegant, and provides the canonical setup in deformation models (see Subsection~\ref{subsec:generativeModel}).  One way of giving a meaning to the notion of expectation in general metric space is to consider the \emph{Fr\'echet mean} (better known in analysis as \emph{barycentre}), named after \cite{frechet1948elements} and defined as the minimiser of the \emph{Fr\'echet functional}
\begin{center}
$
F(\mu)
=\E W^2_2(\Lambda,\mu)
=\ownint{\W_2}{}{W_2^2(\lambda,\mu)}{\P(\lambda)}
,\qquad
\mu\in \W_2,
$
\end{center}
where $\P$ is the law of $\Lambda$.  We shall refer to such a minimiser as the \emph{population (Fr\'echet) mean} to distinguish it from the empirical version, where $\E W^2_2(\Lambda,\mu)$ is replaced with $\sum W_2^2(\mu_i,\mu)$.

Existence, uniqueness, computation, laws of large numbers and central limit theorems for Fr\'echet means with respect to general metrics have been studied extensively, under the umbrella of \emph{non-Euclidean statistics} \citep[e.g.,][]{huckemann2010intrinsic,kendall2011limit}.  Even existence and uniqueness are nontrivial questions for many metrics and depend subtly on the induced geometry.  It turns out that $\W_2$ induces a geometry that is very close to Riemannian (see Subsection~\ref{subsec:geometry}).  Despite posing challenges in that it is infinite-dimensional, has unbounded curvature, and presents an abundance of singularities, its geometry exhibits many favourable (indeed quite unusual for nonlinear spaces) properties owing to the structure of the optimal transport problem.

By means of convex analysis, \cite{agueh2011barycenters} deduce existence,  uniqueness, and a characterisation of empirical Fr\'echet means in $\W_2(\R^d)$, in what has become a seminal paper.  Existence always holds, whereas the mean is unique provided that one of the measures $\mu_i$ is absolutely continuous.  The results extend to the population version \citep{pass2013optimal}:  the condition is that with positive probability $\Lambda$ is absolutely continuous (assuming that the Fr\'echet functional $F$ is finite).  A notable exception is again when $d=1$, in which case Fr\'echet means are unique with the sole restriction that $F$ is finite.

A law of large numbers in Wasserstein space was proved by \cite{legouic2017existence}, in a very general setting (for arbitrary $p>1$, and for spaces more general than $\R^d$).  Since $\W_2(\R^d)$ is itself a complete and separable metric space, one can view $\P$, the law of $\Lambda$, as an element in the ``second level" Wasserstein space $\W_2(\W_2(\R^d))$.  Le Gouic \& Loubes show that if $\P_n$ is a sequence of laws converging to $\P$ in the second level Wasserstein space, then the Fr\'echet means of $\P_n$ converge to that of $\P$ (if unique) in the ``first level" $\W_2(\R^d)$.  This setup covers the case where $\P_n$ is the empirical measure (in $\W_2(\W_2(\R^d))$) corresponding to a sample from $\Lambda$.  See \cite{alvarez2018wide} for an extension to \emph{trimmed} Fr\'echet means.

\subsection{Fr\'echet Means and Generative Models}\label{subsec:generativeModel}
From a statistical perspective, the choice of a metric and the consideration of the corresponding Fr\'echet mean often implicitly assumes a certain underlying data-generating mechanism for the data. In the case of the Wasserstein metric, this mechanism is inextricably linked to \emph{warping} or \emph{phase variation} \citep{ramsay2005functional,marron2015functional,wang2016functional}, where one wishes to infer the law of a process $Y$ on (say) $[0,1]$, but only has access to realisations of $\tilde{Y}=Y\circ T^{-1}$, where $T:[0,1]\to [0,1]$ is a random \emph{warp/deformation} map.  This setup is quite natural in physiological data such as growth curves or spike trains where each individual may have an intrinsic time scale, a sort of functional random effect.  The problem would then be to correct for the effect of $T$ that distorts time, and recover the sample paths in the ``correct", or ``objective" time scale. Typically, it is natural to assume that $T$ is an increasing homeomorphism, on the basis that time should always move forward, rather than backward, and, for identifiability reasons, that $\E T(t)=t,\, t\in[0,1]$.

Now, when the functional datum $Y$ is a random probability measure in $\W_2(\R^d)$ with intensity $\mathbb{E}[Y]=\lambda$, the warped version $\tilde{Y}=T\# Y$ is a random measure with conditional intensity $\Lambda=\mathbb{E}[\tilde{Y}|T]=T\#\lambda$. Assuming that $T$ is increasing with $\E T$ equal to the identity then implies that $\lambda$ is a Fr\'echet mean of $\Lambda$.  More generally, if $\lambda\in\W_2(\R^d)$ and $T$ is a random continuous function with mean identity that can be written as the gradient of a convex function on $\R^d$, then $\lambda$ is a Fr\'echet mean of the random measure $\Lambda=T\#\lambda$.  In other words, the Wasserstein geometry is canonical under the deformation model, and estimation of a Fr\'echet mean implicitly assumes a deformation model.  The result in this form is due to \cite{zemel2017frechet}, but a parametric version is due to \cite{bigot2012characterization}.  When $\lambda$ is absolutely continuous, and $T$ is sufficiently injective, $\Lambda=T\#\lambda$ is absolutely continuous and the Fr\'echet mean of $\Lambda$ is unique, and equals $\lambda$.  In the particular case of Gaussian measures, the result even holds in infinite dimensions \citep{masarotto2018procrustes}.

\subsection{Fr\'echet Means and Multicouplings}\label{sec:frechetMulticoupling}
The Fr\'echet mean problem is related to a \emph{multimarginal} formulation of optimal transport considered by \cite{gangbo1998optimal}.  Given $\mu_1,\dots,\mu_n\in \W_2(\R^d)$, an optimal \emph{multicoupling} is a joint distribution of a random vector $(X_1,\dots,X_n)$ such that  $X_i\sim \mu_i$ and 
\[
\frac1{2n^2}\E \sum_{1\le i<j\le n} \|X_i - X_j\|^2
=\frac 1{2n}\E \sum_{i=1}^n \|X_i - \overline X\|^2,
\]
is minimised.  \cite{agueh2011barycenters} show that if $(X_1,\dots,X_n)$ is an optimal multicoupling, then the law of $\overline{X}=n^{-1}\sum_i X_i$ a is a Fr\'echet mean of $\{\mu_i\}_{i=1}^n$. Inspection of their argument shows that it can also give the ``only if direction". And, when at least one measure $\mu_i$ is regular, necessity and sufficiency combined can be used to construct the optimal multicoupling as $X_i=\mathbf{t}_{\lambda}^{\mu_i}(Z)$, where $Z\sim \lambda$ and $\lambda$ is the Fr\'echet mean (see \cite{pass2013optimal} and \cite{zemel2017frechet} for more details).  This illustrates how constructing the optimal multicoupling is inextricably linked to finding the Fr\'echet mean (for the latter, see Section \ref{steepest_descent}).  In fact, the argument of \cite{agueh2011barycenters} extends to infinite-dimensional and even non-linear space.  Let $(\X,\rho)$ be a complete separable ``barycentric metric space":  for any $x_1,\dots,x_n\in\X$ there exists a unique Fr\'echet mean $\overline{x}$.  Fr\'echet means of given measures $\mu_1,\dots,\mu_n\in \W_2(\X)$ are precisely the laws of $\overline X$, where $(X_1,\dots,X_n)$ is an optimal multicoupling with respect to the cost $\E \sum_{i=1}^n \rho(X_i,\overline X)^2$.  This relation illustrates the idea that the Wasserstein space captures the geometry of the underlying space.  As a particular special case, the Fr\'echet mean of Dirac measures is a Dirac measure at the Fr\'echet mean of the underlying points.  Finally, we stress that the relation extends to any $p>1$, where $\overline x^{(p)}$ minimises $\sum \rho(x_i,x)^p$ and optimality is with respect to $\E \sum \rho(X_i,\overline{X}^{(p)})^p$. (Strictly speaking, these are not Fr\'echet means, as one minimises $\sum W_p^p(\mu_i,\mu)$ instead of $\sum W_p^2(\mu_i,\mu)$.)

\subsection{Geometry of Wasserstein space}\label{subsec:geometry}
A typical step in estimating Fr\'echet means in non-Euclidean settings is approximation of the manifold by a linear space, the tangent space.  In the Wasserstein case, the latter is a function space.  Let $\lambda$ be the Fr\'echet mean, and assume sufficient regularity that $\lambda$ is unique and absolutely continuous.  Then convergence of a sample Fr\'echet mean $\widehat{\lambda}_n$ to $\lambda$ can be quantified by that of the optimal map $\topt \lambda{\widehat{\lambda}_n}$ to the identity map $\id$, because
\[
W_2^2(\widehat{\lambda}_n,\lambda)
=\ownint{\R^d}{}{\|\topt\lambda{\widehat{\lambda}_n}(x) - x\|^2}{\lambda(x)}
=\|\topt\lambda{\widehat{\lambda}_n} - \id\|^2_{\mathcal L^2(\lambda)}.
\]
Here $\mathcal L^2(\lambda)$ is the $L^2$-like space of measurable functions $\mathbf r:\R^d\to\R^d$ such that the real-valued function $x\mapsto \|\mathbf r(x)\|$ is in $L^2(\lambda)$, and whose $L^2(\lambda)$-norm defines $\|\mathbf r\|_{\mathcal L^2(\lambda)}$.  Thus, we can linearise the Wasserstein space by identifying an arbitrary measure $\mu$ with the function $\topt\lambda\mu-\id$ in the linear space $\mathcal L_2(\lambda)$; subtracting the identity ``centres" this linear space at $\lambda$.

\subsubsection{The Tangent Bundle}
\cite{ambrosio2008gradient} consider absolutely continuous curves in Wasserstein space, and show that optimal maps arise as minimial tangent vectors to such curves.  With that in mind they define the tangent space at $\lambda$ as the span of such maps minus the identity:
\[
\Tan_\lambda
=\overline{\{t(\topt\lambda\mu-\id):\mu\in \W_2; t\in\R\}}^{\mathcal L^2(\lambda)}.
\]
By definition, each $\topt\lambda\mu$ (and the identity) is in $\mathcal L^2(\lambda)$, so $\Tan_\lambda\subseteq \mathcal L^2(\lambda)$, from which it inherits the inner product.  The definition can be adapted to a non-absolutely continuous $\lambda$ by restricting $\mu$ in the definition of $\Tan_\lambda$ to those $\mu$ for which $\topt\lambda\mu$ exists (this optimal map might not be unique, and any possible choice of $\topt\lambda\mu$ is in the tangent space).  There is an alternative equivalent definition of the tangent space in terms of gradients of smooth functions, see \citet[Definition~8.4.1 and Theorem~8.5.1]{ambrosio2008gradient}.  The alternative definition highlights that it is essentially the \emph{inner product} that depends on $\lambda$, but not the elements of the tangent space.

The exponential map ${\exp}_\lambda:\Tan_\lambda \to \W_2$ at $\lambda$ is the restriction of the transformation that sends $\mathbf r\in \mathcal L^2(\lambda)$ to $(\mathbf r + \id)\#\lambda\in \W_2$.  Specifically,
\[
{\exp}_{\lambda}(t(\mathbf t - \id))
=[t(\mathbf t - \id) + \id)\#\lambda
= [t\mathbf t + (1-t)\id]\#\lambda
\quad(t\in\R).
\]
When $\lambda$ is absolutely continuous, the log map ${\log}_\lambda:\W_2 \to \Tan_\lambda$ is
\[
\log_{\lambda}(\mu)
=\topt{\lambda}{\mu} - \id,
\]
and is the right inverse of the exponential map (which is therefore surjective).  Segments in the tangent space are retracted to the Wasserstein space under $\exp_\lambda$ to McCann's (1997) interpolant \phantom{\cite{mccann1997convexity}}
\[
\left[t(\topt{\lambda}{\mu} + (1-t)\id\right]\#\lambda,
\]
and these are the unique (constant speed) geodesics in Wasserstein space \citep[Proposition 5.32]{santambrogio2015optimal}.  If $\lambda$ is singular, then the log map is only defined on a subset of Wasserstein space.  See \cite{gigli2011inverse} for a description of the tangent bundle when the underlying space $\R^d$ is replaced by a Riemannian manifold.

\subsubsection{Curvature and Compatible Measures}
If $\mu,\nu,\rho\in\W_2$, then a coupling argument shows that
\begin{equation}\label{eq:positiveCurvature}
\|\log_\rho(\mu) - \log_\rho(\nu)\|_{\mathcal L^2(\rho)}^2
=
\|\topt \rho\mu - \topt\rho\nu\|_{\mathcal L^2(\rho)}^2
=\ownint {}{}{\|\topt\rho\mu(x) - \topt\rho\nu(x)\|^2}{\rho(x)}
\ge W_2^2(\mu,\nu).
\end{equation}
In differential geometry terminology, this means that $\W_2$ has nonnegative sectional curvature.  In the special case $d=1$, there is equality, and the Wasserstein space is flat;  the correspondence $\mu\iff \topt\rho\mu-\id$ is an \emph{isometry}, and $\W_2(\R)$ can be viewed as a subset of the Hilbert space $L^2(\mu)$.  Computation of Fr\'echet means in then particularly simple:  if $\mu_1,\dots,\mu_n$ are arbitrary measures in $\W_2(\R)$ and $\nu$ is any absolutely continuous measure, then the Fr\'echet mean of $(\mu_i)$ is $[(1/n)\sum \topt\nu{\mu_i}]\#\nu$;  this extends to the population version.  An important extension to $\R^d$ was obtained by \cite{boissard2015distribution}.  Equality will hold in Equation~\ref{eq:positiveCurvature} provided some ``compatibility" holds between the measures $\mu,\nu,\rho$.  The composition $\topt\rho\nu\circ \topt\mu\rho$ pushes $\mu$ forward to $\rho$ by definition, but might not be the optimal one.  We say that $\mu,\nu,\rho$ are \emph{compatible} if $\topt\rho\nu\circ \topt\mu\rho$ is optimal, i.e., equals $\topt\mu\nu$.  \cite{boissard2015distribution} show that if the collection $(\mu_1,\dots,\mu_n,\nu)$ is compatible (in their terminology, the optimal maps are \emph{admissible}) in this sense, then again the Fr\'echet mean is $[(1/n)\sum \topt\nu{\mu_i}]\#\nu$.  This setup covers the one-dimensional setup, but also multivariate measures with structure that mimics the one-dimensional case.  For example, a collection of measures having the same $d$-dimensional copula (and potentially different marginals) is compatible, and so is a collection of measures having the same ``angular" behaviour but different marginal distributions for their norms.

\subsubsection{Gaussian Measures}
Without such structural restrictions the Wasserstein space is positively curved, and computation of the Fr\'echet mean of a sample is not straightforward.  As an important example, if $\mu_i\sim N(0,\Sigma_i)$ are nonsingular Gaussian measures on $\R^d$, then the Fr\'echet mean is also Gaussian and its covariance is the unique nonsingular solution of the matrix equation
\begin{equation}\label{eq:GaussFrechet}
\Sigma
=\frac 1n \sum_{i=1}^n (\Sigma^{1/2}\Sigma_i\Sigma^{1/2})^{1/2}.
\end{equation}
The $\mu_i$'s will be compatible if the covariances commute, in which case we have the explicit solution $\Sigma^{1/2}=n^{-1}(\Sigma_1^{1/2}+\dots+\Sigma_n^{1/2})$, but otherwise there is no explicit expression for the Fr\'echet mean.  The restriction of $\W_2(\R^d)$ to Gaussian measures leads to a \emph{stratified space}, whose geometry was studied carefully by \cite{takatsu2011wasserstein}, including expressions for the curvature.  In particular, the latter grows without bound as one approaches singular covariance matrices.

\subsection{Fr\'echet Means via Steepest Descent}\label{steepest_descent}
A common procedure for finding Fr\'echet means is differentiation of the Fr\'echet functional $F$ and moving in the negative direction of the gradient \citep{karcher1977riemannian,afsari2013convergence}.  The gradient at $x_0$ typically takes the form
\[
\nabla F(x)
=\frac1n \sum_{i=1}^n-\log_{x}(x_i).
\]
This formula also holds true in Wasserstein space, where the log map is as given in Subsection~\ref{subsec:geometry}.  Steepest descent can then be defined using the exponential map as
\[
\rho_{j+1}
=\exp_{\rho_j}(\nabla F(\rho_j))
=\left[\frac 1n\sum_{i=1}^n\topt{\rho_j}{\mu_i}\right]\#\rho_j.
\]
The resulting iteration was independently arrived at in this steepest descent form by \cite{zemel2017frechet} and in the form of a fixed point equation iteration by \cite{alvarez2016fixed}. It has the advantage of reducing the multitransport problem of finding the Fr\'echet mean to a succession of pairwise problems that are simpler in nature, in the same spirit as \emph{generalised Procrustes analysis} \citep{dryden1998statistical}.  This benefit is best illustrated in the Gaussian case, where the optimal maps have the explicit expression given in Equation~\ref{eq:gaussTransport}.  The algorithm converges to the unique Fr\'echet mean in this Gaussian case, and in general will reach at least a stationary point (where $\nabla F$ vanishes).  There are local minima that are not global: \cite{alvarez2016fixed} construct measures $\mu_1,\dots,\mu_4,\mu$ in $\R^2$ such that the average of $\topt\mu{\mu_i}$ is the identity, but $\mu$ is not the Fr\'echet mean.  Their example shows that the problem cannot be solved by smoothness conditions on the measures.  But smoothness and convexity of the supports yields an optimality criterion for local minima \citep{zemel2017frechet}, roughly in that a sufficiently smooth local minimum is a global minimum.

\subsection{Large Sample Statistical Theory in Wasserstein Space}
The general consistency result of \cite{legouic2017existence} is the important and necessary first step in providing a sound statistical theory for random measures in Wasserstein space.  The next step would be to establish the rate of convergence and a central limit theorem.  Exploiting the central limit theorem in Hilbert spaces, the one-dimensional case is well-understood, even under sampling noise: the empirical mean $\widehat{\lambda}_n$, viewed as the $L^2$ map, $\sqrt n(\topt\lambda{\widehat{\lambda}_n} - \id)$, converges in distribution to a zero-mean Gaussian process whose covariance structure is that of the random element $\topt\lambda\Lambda$ \citep{panaretos2016amplitude};  see \cite{bigot2018upper} for minimax-type results in this vein.  Since the Wasserstein space on $\R^d$ stays embedded in a Hilbert under the compatible setup of \cite{boissard2015distribution}, these results can certainly be extended to that setup.  In fact, \cite{boissard2015distribution} use this embedding to carry out principal component analysis (PCA) in Wasserstein space.  See \cite{bigot2017geodesic} for an alternative procedure, \emph{convex PCA}.

The only central limit theorem-type result we know of beyond the compatible setup was found recently by \cite{agueh2017vers}.  Suppose that $\Lambda$ takes finitely many values: $\P(\Lambda=\lambda_k)=p_k$, $k=1,\dots,K$, and $\lambda_k$ is Gaussian $N(0,\Sigma_k)$ with $\Sigma_k$ nonsingular.  Given a sample $\mu_1,\dots,\mu_n$ from $\Lambda$, let $\widehat{p}_n(k)$ be the proportion of $(\mu_i)$'s that equal $\lambda_k$.  Then $\sqrt n(\widehat{p}_n - p)$ has a Gaussian limit.  Equation~\ref{eq:GaussFrechet} extends to weighted Fr\'echet means, and defines $\Sigma$ in a sufficiently smooth way, so one can invoke the delta method to obtain a central limit theorm for $\sqrt n(\widehat{\Sigma}_n-\Sigma)$.  \cite{agueh2017vers} also cover the case $K=2$ and $\lambda_i$ arbitrary, though this setup falls under the umbrella of compatibility, since any pair of measures is a compatible collection.  Ongoing work by \cite{kroshnin2018central} focusses on extending the results of \cite{agueh2017vers} to arbitrary random Gaussian/elliptical measures.  Beyond this location-scatter setup, very recent results  by \cite{ahidar2018rate} suggest that the rate of convergence of the empirical Fr\'echet mean to its population counterpart can be slower than $n^{-1/2}$.

\section{Computational Aspects}\label{sec:numerics}
Beyond the one-dimensional and Gaussian cases, explicit expressions for the Wasserstein distance and/or the optimal couplings are rare.  When $\mu=(1/n)\sum_{i=1}^n\delta_{x_i}$ and $\nu=(1/m)\sum_{j=1}^m\delta_{y_j}$ are uniform discrete measures on $n$ and $m$ points, a coupling $\gamma$ can be identified with an $n\times m$ matrix $\Gamma$, where $\Gamma_{ij}$ represents the mass to be transferred from $x_i$ to $y_j$.  The cost function reduces to a cost matrix $c_{ij}=\|x_i - y_j\|^p$, and the total cost associated with it is $\sum_{ij}\Gamma_{ij}c_{ij}$.  This double sum is to be minimised over $\Gamma$ subject to the $m+n$ mass preservation constraints
\[
\sum_{i=1}^n \Gamma_{ij} = 1/m \quad (j=1,\dots,m),
\qquad
\sum_{j=1}^m \Gamma_{ij} = 1/n \quad (i=1,\dots,n),
\qquad \Gamma_{ij}\ge0.
\]
One can easily write the constraints in the weighted version of the problem.  This optimisation problem can be solved using standard linear programming techniques.  In particular, there exists an optimal solution $\Gamma$ with at most $n+m-1$ nonzero entries.  In the special case $n=m$ and uniform measures, the extremal points of the constraints polytope are the permutation matrices, and these correspond precisely to deterministic couplings, that have $n$ (rather than $2n-1$) nonzero entries.

The specific structure of the constraints matrix allows the development of specialised algorithms:  the Hungarian method of \cite{kuhn1955hungarian} and its variant by \cite{munkres1957algorithms} are classical examples, with alternatives such as network simplex, min flow-type algorithms and others \citep[see][Chapter~6]{luenberger2008linear}.  The best algorithms have the prohibitive complexity $n^3\log n$ in the worst-case scenario.  \cite{sommerfeld2018optimal} propose sampling $s<<n$ points from $\mu$ and $\nu$ and estimating $W_p(\mu,\nu)$ by the empirical distance $W_p(\mu_s,\nu_s)$.  They provide bounds on the computational and statistical trade-off regulated by $s$.

The multimarginal problem can also be recast as a linear program whose solution yields the Fr\'echet mean (see Subsection \ref{sec:frechetMulticoupling}).  If we have $n$ measures $\mu_i$ supported on $m_i$ points ($i=1,\dots,n$), then the number of variables in the problem is $\prod m_i$, and the number of equality constraints is $\sum m_i$, of which $n-1$ are redundant.  See \cite{anderes2016discrete} for a detailed account of the problem, where they show the peculiar property that the optimal maps $\topt{\overline \mu}{\mu_i}$ exist, where $\overline \mu$ is a Fr\'echet mean.  This is far from obvious, since besides the uniform discrete setup with equal number of points, the optimal coupling between discrete measures is rarely induced from a map. There are alternative formulations with fewer variables and fewer constraints:  exact ones \citep{borgwardt2018improved} as well as polynomial-time approximations \citep{borgwardt2017strongly}.

One can certainly approximate $W_p(\mu,\nu)$ by $W_p(\mu_n,\nu_n)$ for some $\mu_n,\nu_n$ supported on, say, $n$ points.  The approximated problem can be solved exactly, as it is a finite linear program.  How to best approximate a measure by discrete measures amounts to \emph{quantisation} and is treated in detail in \cite{graf2007foundations}.  Unfortunately, quantisation is extremely difficult in practice, and even one-dimensional measures rarely admit explicit solutions, and, moreover, the computational cost of solving the $n$-to-$n$ points scales badly with $n$.

Another class of algorithm is ``continuous" in nature.  Recall from Subsection~\ref{subsec:geometry} that optimal maps $\topt\mu\nu$ are equivalent to the unique geodesics in $\W_2$.   \cite{benamou2000computational} exploit this equivalence and develop a numerical scheme to approximate the entire geodesic.  Although this dynamic formulation adds an extra ``time" dimension to the problem, it can be recast as a convex problem, unlike the formulation with the optimal map as variable.  \cite{chartrand2009gradient} carry out steepest descent in the dual variable $\varphi$ in order to maximise the dual $\varphi\mapsto \ownint{}{}\varphi\mu+\ownint{}{}{\varphi^*}\nu$.

In an influential paper, \cite{cuturi2013sinkhorn} advocated adding an \emph{entropy} penalty term $\kappa \sum \Gamma_{ij}\log \Gamma_{ij}$ to the objective function.  This yields a strictly convex problem with complexity $n^2$, much smaller than the linear programming complexity $n^3\log n$.  This entropy term enforces $\Gamma$ to be diffuse (strictly positive), in stark contrast with the unpenalised optimal coupling, but the regularised solution converges to the sparse one as $\kappa\searrow 0$.  This idea is extended to the Fr\'echet mean problem in \cite{cuturi2014fast}, where the Fr\'echet mean is computed with respect to the penalised Wasserstein distance, and in \cite{bigot2017penalized}, where the penalisation is imposed on the mean itself, rather than the distance.  \cite{bigot2018data} suggest a data-driven choice of the regularisation parameter according to the Goldenshluger--Lepski principle.

This field of research is very active, and there are tens of extensions and new algorithms.  One can find a short survey in \cite{tameling2018computational}, and we refer to \citet[Chapter 6]{santambrogio2015optimal} and especially the forthcoming book \cite{peyre2018computational} for more details and references.

\section{On Some Related Developments}\label{sec:MoreReferences}
An interesting recent development that is, strictly speaking, not so much about Wasserstein distances, as about measure transportation itself, considers how to generalise notions related to quantiles to several dimensions. In one dimension, the quantile function $F_Y^{-1}$ is the optimal map from a uniform variable $U$ to $Y$.  This observation can be used in order to define a multivariate quantile function of $Y$ using the optimal transport map $\topt UY$ from some reference random variable $U$ (e.g., uniform on the unit ball).  \cite{chernozhukov2017monge} describe the resulting form of the quantile contours and the induced notions of depth and ranks, and estimate them from data.  Further work by \cite{hallin2017distribution} considers extensions of the approach that do not require finite variance for $Y$ (as is the case in one dimension).  This measure-transportation approach also allows to extend quantile regression to multivariate setups \citep{carlier2016vector}.

Finally, due to space considerations we have not attempted to describe the machine learning side of optimal transport, though there is a fast-growing literature for such tasks. Indicative examples include estimation of a low-dimensional measure in high-dimensional space \citep{canas2012learning}, regression in the space of histograms \citep{bonneel2016wasserstein}, dictionary learning \citep{rolet2016fast}, Gaussian processes indexed by measures on $\R$ \citep{bachoc2017gaussian} or $\R^d$ \citep{bachoc2018gaussian}, clustering in Wasserstein space \citep{delBarrio2018robust}, and unsupervised alignment of point clouds in high dimensions \citep{grave2018unsupervised}.

\section*{Acknowledgments}
This was supported in part by an European Reasearch Council Starting Grant Award to Victor M. Panaretos. Yoav Zemel is funded by Swiss National Science Foundation grant \#178220.  We thank a reviewer for comments on a preliminary version of the paper.

\end{document}